\documentclass[preprint,3p,12pt,authoryear,times]{elsarticle}
\usepackage[english]{babel}
\usepackage{amsmath, amsthm, amssymb, latexsym, graphicx}
\usepackage[]{natbib}
\usepackage[utf8]{inputenc}

\begin{document}
%\linenumbers

\begin{frontmatter}
\title{Saturn's ULF wave foreshock boundary: Cassini observations}
 
\author[iafe]{N. Andr\'es}
\ead{n\_andres@iafe.uba.ar}
\cortext[cor1]{Corresponding author. Av. Cantilo 2620 Pabell\'on IAFE (cp 1428), CABA, Argentina. Tel.: +5411 47890179 (int. 134) ; Fax: +5411 47868114.}

\author[iafe,df]{D. O. G\'omez}
\ead{dgomez@iafe.uba.ar}
 
\author[iafe]{C. Bertucci}
\ead{cbertucci@iafe.uba.ar}

\author[france]{C. Mazelle}
\ead{christian.mazelle@irap.omp.eu}

\author[london]{M. K. Dougherty}
\ead{m.dougherty@imperial.ac.uk}

\address[iafe]{Instituto de Astronom\'ia y F\'isica del Espacio, Buenos Aires, Argentina}
\address[df]{Departamento de F\'isica, FCEN, UBA, Buenos Aires, Argentina}
\address[france]{Institut de Recherche en Astrophysique et Plan\'etologie, Toulouse, France}
\address[london]{Space and Atmospheric Physics, The Blackett Laboratory, Imperial College, London, United Kingdom}

\begin{abstract}
Even though the solar wind is highly supersonic, intense ultra-low frequency (ULF) wave activity has been detected in regions just upstream of the bow shocks of magnetized planets. This feature was first observed ahead of the Earth's bow shock, and the corresponding region was called the ULF wave foreshock, which is embedded within the planet's foreshock. The properties as well as the spatial distribution of ULF waves within the Earth's foreshock have been extensively studied over the last three decades and have been explained as a result of plasma instabilities triggered by solar wind ions backstreaming from the bow shock. Since July 2004, the Cassini spacecraft has characterized the Saturnian plasma environment including its upstream region. Since Cassini's Saturn orbit insertion (SOI) in June 2004 through August 2005, we conducted a detailed survey and analysis of observations made by the Vector Helium Magnetometer (VHM). The purpose of the present study is to characterize the properties of waves observed 
in Saturn's ULF wave foreshock and identify its boundary using single spacecraft techniques. The amplitude of these waves is usually comparable to the mean magnetic field intensity, while their frequencies in the spacecraft frame yields two clearly differentiated types of waves: one with frequencies below the local proton cyclotron frequency ($\Omega_{\text{H+}}$) and another with frequencies above $\Omega_{\text{H+}}$. All the wave crossings described here, clearly show that these waves are associated to Saturn's foreshock. In particular, the presence of waves is associated with the change in $\theta_{Bn}$ to quasi-parallel geometries. Our results show the existence of a clear boundary for Saturn's ULF wave foreshock, compatible with $\theta_{Bn}\sim45^{\circ}$ surfaces.
\end{abstract}

 
\end{frontmatter}
 
\section{Introduction}\label{int}

When the supersonic solar wind plasma from the Sun encounters an obstacle, a bow shock is formed. The incoming solar wind particles (electrons and ions) upstream from the bow shock have no information about the obstacle, except in a region magnetically connected to it. This region is known as the foreshock. At the bow shock, a small fraction of the solar wind particles are reflected in the sunward direction. These backstreaming particles are subjected to the solar wind's \textbf{E}$\times$\textbf{B} drift, where \textbf{E}$=-$\textbf{v}$_\text{sw}\times$\textbf{B}$/c$, is the solar wind's convective electric field, \textbf{B} is the interplanetary magnetic field (IMF), $\textbf{v}_\text{sw}$ is the solar wind velocity and $c$ is the speed of light. The \textbf{E}$\times$\textbf{B} drift velocity is the same for all backstreaming particles, and perpendicular to the IMF. As a result, the guiding centers of all backstreaming particles move within the $\textbf{v}_\text{sw}$-\textbf{B} plane, gradually drifting 
away from the field line tangent to the bow shock toward the inner part of the foreshock and being segregated according to their parallel velocities. Electrons, because of their much smaller inertia, are much less affected by this drift and their presence can be detected right next to the field line tangent to the bow shock. The combination of solar wind and energetic backstreaming electrons results in the production of Langmuir waves at the electron plasma frequency \citep{F1985,S2004a,S2004b}. Backstreaming ions on the other hand, can drive a number of plasma instabilities \citep{G1993,CG1997}, leading to the generation of waves. The ion foreshock is then characterized not just by the presence of a small fraction of backstreaming ions, but also by the generation and propagation of plasma waves around the local ion cyclotron frequency.

The understanding of planetary foreshocks is far from complete. The most studied case is the Earth's foreshock, which is reasonably well understood thanks to single and multi-spacecraft measurements \citep{TR1981,E2005}. The first observations from the Earth's foreshock were made by the dual spacecraft ISEE \citep{O1977} which identified different types of backstreaming ion distributions: reflected (now called field-aligned beams), intermediate, and diffuse \citep{G1978,P1981}. The classification of backstreaming ion populations into these three types was made on the basis of two-dimensional velocity distribution functions and energy-time spectrograms. Further results from ISEE demonstrated the existence of gy\-ro\-pha\-se-bun\-ched and gyrotropic backstreaming ion distributions in the foreshock \citep{G1983}. 
%Since its launch in 2000, observations from the multi-spacecraft mission Cluster \citep{B1997} have also been used to understand the foreshock morphology \citep{E2005}.

The field-aligned distributions are typically observed at and near the leading edge of the ion foreshock without the presence of waves \citep{P1979}. Behind the field-aligned beam region, gyrophase-bunched distributions are detected, while diffuse distributions are found even farther away (i.e. downstream) from the ion foreshock boundary. The association of different linear and non-linear waves to different ion distributions was first studied by \citet{HR1983}. Gy\-ro\-pha\-se-bun\-ched and diffuse distributions are in fact observed in the presence of ultra-low frequency (ULF) waves. In particular, gyrophase-bunched distributions coexist with ULF quasi-monochromatic waves with substantial amplitudes ($\delta\textbf{B}/B\approx1$) \citep{M2003}. The production of gyrophase-bunched ions and associated ULF waves has been studied numerically \citep{HT1985}, theoretically \citep{M2000} and observationally, using data from WIND spacecraft \citep{M2001}. On the other hand, non-linear, steepened waves have been 
found to be associated with diffuse ion distributions \citep{H1981}. 

Evidence of foreshocks has also been found in other planets as well. In the case of Mars, \citet{T1992} found electrostatic waves which are polarized along the interplanetary magnetic field, and their peak intensity occurs at or near the local solar wind plasma frequency. \citet{G1987} identified the ULF wave foreshock boundary of Venus through observations made by the PVO (Pioneer Venus Orbiter) magnetometer. Plasma and magnetic field observations from the Voyager 2 spacecraft reveal ULF waves in the solar wind, which are associated with Neptune's, Jupiter's and Uranus' foreshocks \citep[see for example][]{Be1991,B1987,R1990}. Plasma waves have also been detected upstream from Mercury's bow shock using Mariner 10 measurements \citep{FB1976}. A brief review on Mercury's foreshock was made by \citet{Bl2007}.

The region of ULF wave activity is embedded in the ion foreshock, and its boundary is known as the ULF wave foreshock boundary. In the case of the Earth, \citet{D1976} located this boundary in a statistical way by using combined magnetic field and plasma data from Heos 1. \citet{GB1986}, using ISSE 1 data introduced the so called solar foreshock coordinates, and determined the position of the ULF wave foreshock boundary. For different IMF cone angles (they split their study into $\theta_{Bx}=20^{\circ}$ - $30^{\circ}$ and $\theta_{Bx}=40^\circ$ - $50^\circ$ data sets), \citet{GB1986} obtained different orientations of the ULF wave foreshock boundary. Other studies have shown that for $\theta_{Bx}\geq45^\circ$ cone angles, there is a well defined upstream boundary, and this boundary intersects the bow shock at $\theta_{Bn}\approx50^\circ$ \citep{LR1992a}. Using multi-spacecraft data obtained by Cluster, \citet{E2005} presented two case studies on the onset of ULF foreshock waves, relating their appearance to 
changes in the orientation 
of the IMF. The increased wave activity is associated with the change in $\theta_{Bn}$ to quasi-parallel geometries ($\theta_{Bn}<45^\circ$). These findings are consistent with previous single spacecraft studies.

Although the observations at Earth represent a vital element in the study of the physical processes occurring on planetary foreshocks at large, these phenomena necessarily occur at the particular parameter values relevant for our planet. To explore how these processes change in parameter space, it is just as important to make in situ observations around other planets. For instance, the solar wind properties vary with heliocentric distance. The Parker angle between the stream lines and the radial direction to the Sun at 1 AU is predicted to be about 45$^\circ$ for a constant solar wind velocity of 429 km$/$s \citep{TS1980}. At Saturn's distance (approx. 9 AU), the Parker angle is predicted to be about 85$^\circ$. \citet{J2008} examined in detail the hourly averaged IMF data provided by Cassini from 13 August 2003 to 14 November 2004, and found an average Parker angle equal to 86.8$^{\circ}\pm$0.3$^\circ$ for a solar wind speed of 500 km$/$s.

The location and shape of planetary bow shocks are determined by the properties of the solar wind flow and by the size and shape of the obstacle to the flow. Saturn's bow shock was first observed by Pioneer 11 in 1979 \citep{A1980}. The first crossings made by Cassini during Saturn Orbit Insertion (SOI) were analyzed and discussed by \citet{A2006}. They presented evidence of magnetospheric compression during Cassini's first immersion into the magnetosphere and the properties of the solar wind upstream from Saturn's bow shock for the first six bow shock crossings observed by Cassini. The magnetic signatures of these bow shock crossings showed a clearly defined overshoot and foot regions associated with the quasi-perpendicular geometry ($\theta_{Bn}>45^\circ$). Using magnetic field and plasma observations made by Cassini between June 2004 and August 2005, \citet{M2008} presented a static model of Saturn's bow shock. The model was obtained by fitting a conic section to the first 206 crossings observed by 
Cassini. Based on observations from Pioneer 11, Voyager 1, Voyager 2 and Cassini \citet{W2011} derive a small eccentricity for Saturn's bow shock and found variations in the shock subsolar distance associated to variations in the solar wind dynamic pressure.

Using the Magnetospheric Imaging Instrument (MIMI) on board Cassini, \citet{K2005} showed measurements of the energetic ion population upstream from both the dusk and dawn sides of the Kronian magnetosphere. During the approach phase and first orbits of Cassini, the observations revealed the presence of a series of distinct upstream bursts of energetic hydrogen and oxygen ions up to distances of 120 Saturn's radii. They concluded that these oxygen upstream events must be particles leaking from Saturn's magnetosphere under favorable IMF conditions. However, \citet{T2007} studied 45 hours of mass-resolved observations of Cassini Plasma Spectrometer (CAPS), which were performed upstream from Saturn's bow shock. The observations show supra-thermal ions composed of H$^+$ and ions with $m/q=2$, presumably solar wind He$^{++}$, with no detectable contribution from magnetospheric water group ions. 

\citet{B1991} reported the first evidence of upstream low-frequency waves in Saturn. During these observations, the spacecraft was magnetically connected to Saturn's bow shock. Their results suggest that these waves were associated to the planet's foreshock. These waves displayed a period of 550 seconds in the spacecraft frame and a relative amplitude of 0.3. Also in the spacecraft frame, the waves are left and right-hand elliptically polarized, and propagate at about 30$^\circ$ with respect to the ambient magnetic field. During the first three orbits of Cassini spacecraft, \citet{B2007} present a characterization of low-frequency waves associated with Saturn's foreshock based on Cassini magnetometer \citep{D2004}. As a result of their survey, they identified two distinct types of waves. They found a large majority of waves with spacecraft-frame frequencies below the local proton cyclotron frequency ($\Omega_{\text{H+}}=\text{eB}_{\text{0}}/\text{m}_{\text{H}}\text{c}$). These waves are phase-steepened and 
display a left-hand elliptical polarization as seen by the spacecraft. \citet{B2007} interpreted these waves as fast magnetosonic waves. This kind of waves is the same presented by \citet{B1991}. In a second group, they found waves with frequencies above $\Omega_{\text{H+}}$, quasi-monochromatic and steepened with a right-hand circular polarization, propagating at small angles with respect to the ambient field. According to \citet{B2007} these waves could be Alfv{\'e}n waves similar to those observed at Earth by \citet{E2003}. 

In this paper, we have used Cassini magnetometer data and single-spacecraft techniques to study the morphology of Saturn's ULF wave foreshock. In situ observations made by the Vector Helium Magnetometer (VHM) on board Cassini are presented in section \ref{obs}. A brief study of ULF waves in the magnetic field associated with Saturn's foreshock is presented in section \ref{ulf-waves}. In section \ref{ulf-bound} we show, for the first time, the determination of the outer boundary of Saturn's ULF wave foreshock. For this purpose we identified a large number of crossings of Cassini to (or from) the wave region. The selection criterion for these crossing is described in section \ref{dtc}. In sections \ref{res} and \ref{theta} we introduced the solar foreshock coordinates developed by \citet{GB1986}, and present our main results. Discussion and our conclusions are summarized in section \ref{dis} and \ref{con}.

\section{Observations}\label{obs}

\begin{figure}[h!]
\begin{center}
\includegraphics[width=1\textwidth]{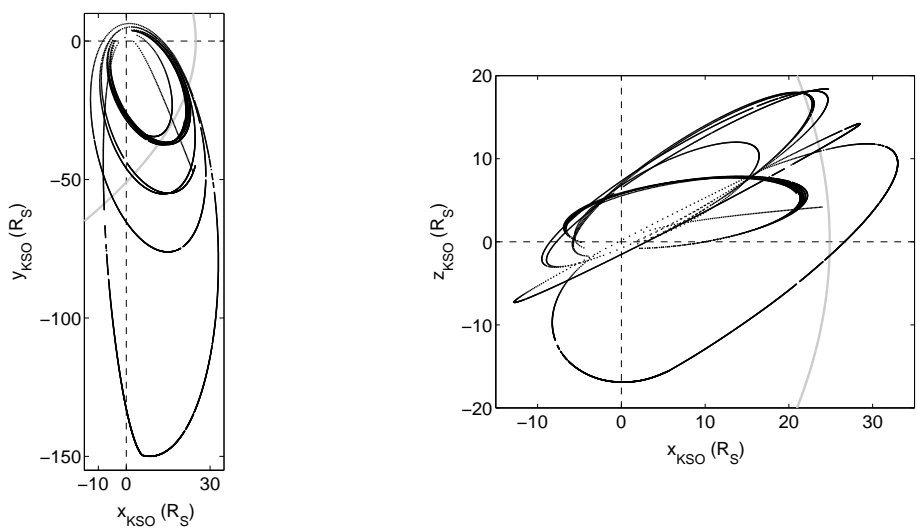}
\end{center}
\caption{Cassini's trajectory, since SOI in June 2004 until August 2005, projected into the ${\textbf{x}}$-${\textbf{y}}$ and the ${\textbf{x}}$-${\textbf{z}}$ plane in Kronocentric Solar Orbital coordinates. The average bow shock fit is the gray-solid line.}
\label{fig:orbita}
\end{figure}

The observations used for the present study consist of the three components of magnetic field measured by Cassini MAG obtained upstream from the Kronian shock during the first fifteen months of Cassini's orbital data, i.e. from SOI in June 2004 through August 2005. Cassini's trajectory in Kronian Solar Orbital (KSO) coordinates for that period is shown in Figure \ref{fig:orbita}. In the KSO coordinate system centered on Saturn, the  $\hat{\textbf{x}}_\text{kso}$ axis points to the Sun, the $\hat{\textbf{y}}_\text{kso}$ axis is anti-parallel to Saturn's orbital velocity and the $\hat{\textbf{z}}_\text{kso}$ axis points towards the north pole of the Ecliptic. During this phase of the orbital tour, the spacecraft explored the dawn side of Saturn's magnetosphere at low Kronographic latitudes. The Cassini dual magnetometer investigation consists of a Vector Helium Magnetometer (VHM) and a Fluxgate Magnetometer (FGM) which provide redundant, fast vector measurements of the ambient magnetic field over a wide range. 
The VHM provides accurate vector measurements with a resolution of 2 s$^{-1}$ over a range of $\pm256$ nT, whereas FGM samples the magnetic field over a larger range ($\pm65655$ nT) and at higher frequency (32 s$^{-1}$). This dual technique is extensively described in \citet{D2004}. For the purpose of this paper, we only consider the data provided by the VHM. Unfortunately, CAPS could not be used in this study due to the absence of periods with adequate pointing. 

In the absence of multi-spacecraft observations, the reconstruction of the foreshock's geometry requires a model of the Kronian bow shock. In this work, we used a static model based on a fit of the crossings using a conic section. The functional form for the fit was introduced by \citet{S1985}, assuming that the bow shock is axially symmetric about the solar wind flow direction. The general equation of a conic section is given by
\begin{equation}\label{conic}
 r=\frac{L}{1+e\text{cos}\theta}
\end{equation}
where $r$ is the distance from the planet to a point on the shock surface, $\theta$ is the corresponding polar coordinate angle with respect to the symmetry axis, $L$ is the semilatus rectum (size parameter) and $e$ is the eccentricity. An empirical model has been presented by \citet{M2008} using measurements from Cassini MAG, CAPS' electron spectrometer sensor (ELS) and density measurements from the Radio Plasma Wave System (RPWS). Assuming a constant solar wind speed ($\approx500$ km/s) they found an eccentricity of $e=1.05\pm0.09$ (hyperboloid) and a size parameter for the average bow shock location of $L=(51\pm2)~R_{S}$, which implies that the average standoff distance is $(25\pm1)~R_{S}$. Throughout their study, they did not find any dependence of the global configuration of the bow shock model with the IMF cone angle.

\section{Characterization of ULF waves}\label{ulf-waves}

\begin{figure}[h!]
\begin{center}
\includegraphics[width=1\textwidth]{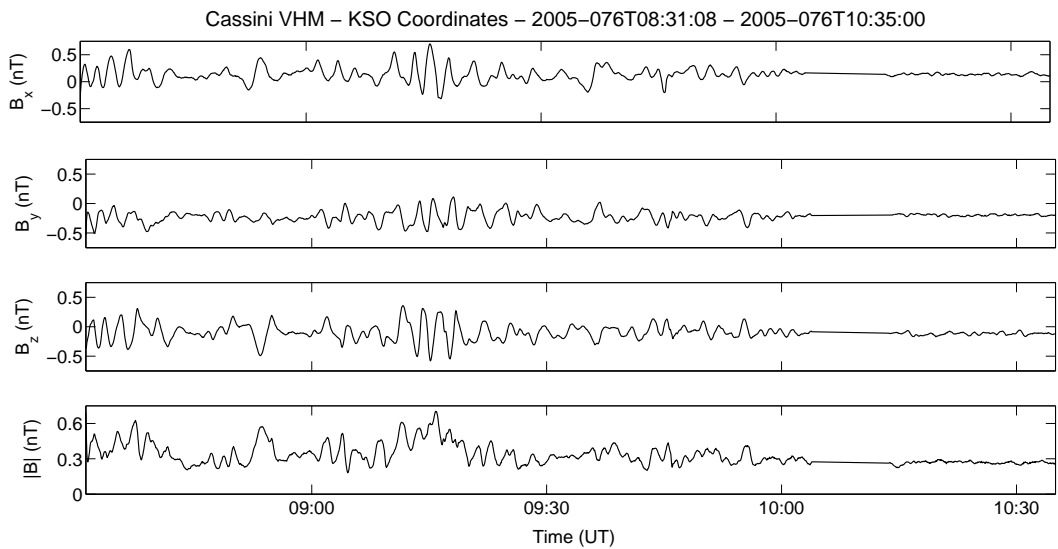}
\end{center}
\caption{Example of a wave train detected by the VHM on board Cassini on 17 March (day 076) 2005 between 08:35:00 UT and 10:35:00 UT. The quasi-monochromatic packet between 08:31:12 UT and 08:34:04 UT is analyzed in Figure \ref{fig:mva1}.}
\label{fig:waves_m}
\end{figure}

\begin{figure}[h!]
\begin{center}
\includegraphics[width=1\textwidth]{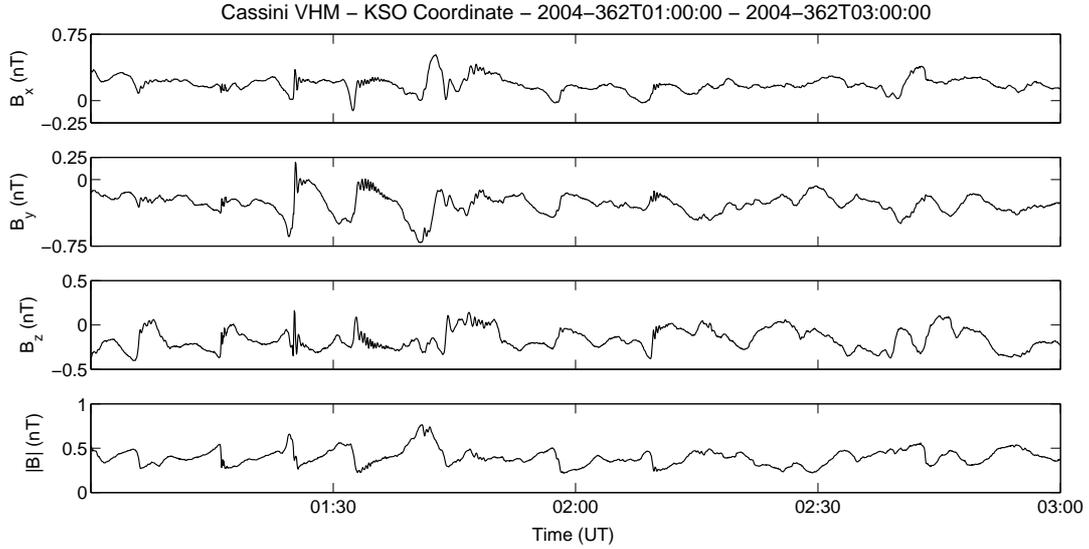}
\end{center}
\caption{Example of a wave train detected by the VHM on board Cassini on 27 December (day 362) 2004 between 01:00:00 UT and 03:00:00 UT. This wave event corresponds to a non linear packet. The packet between 01:30:36 UT and 01:33:00 UT is analyzed in Figure \ref{fig:mva2}.}
\label{fig:waves_nl}
\end{figure}

During Cassini's excursions into the solar wind, we looked for intervals with low-frequency fluctuations in the magnetic field components. We found intervals with and without the presence of waves. Figure \ref{fig:waves_m} and figure \ref{fig:waves_nl} show two examples of the types of wave events found in this paper.

For all the wave events, we made a characterization (polarization and frequency) in the spacecraft frame. We calculated their frequencies $\omega$ (in the spacecraft frame). As a result of this study, we find two distinct types of oscillations with different properties, depending on whether their frequencies are below or above the local proton cyclotron frequency ($\Omega_{\text{H+}}=\text{eB}_{\text{0}}/\text{m}_{\text{H}}\text{c}$). We also studied their polarization and propagation with respect to the ambient magnetic field using the Minimum Variance Analysis (MVA) \citep{SS1998}. The MVA consists in building the variance matrix in terms of the measured magnetic field components for a given time interval and finding the three eigenvalues and corresponding eigenvectors. The eigenvector corresponding to the smallest eigenvalue $\lambda_3$ (the maximum and intermediate eigenvalues are respectively $\lambda_1$ and $\lambda_2$) is used as an estimate of the direction of propagation of a plane wave. 
Note that the eigenvector set of the variance matrix provides a convenient natural coordinate system in which to display and analyse the data. 

\begin{figure}[h!]
\begin{center}
\includegraphics[width=1\textwidth]{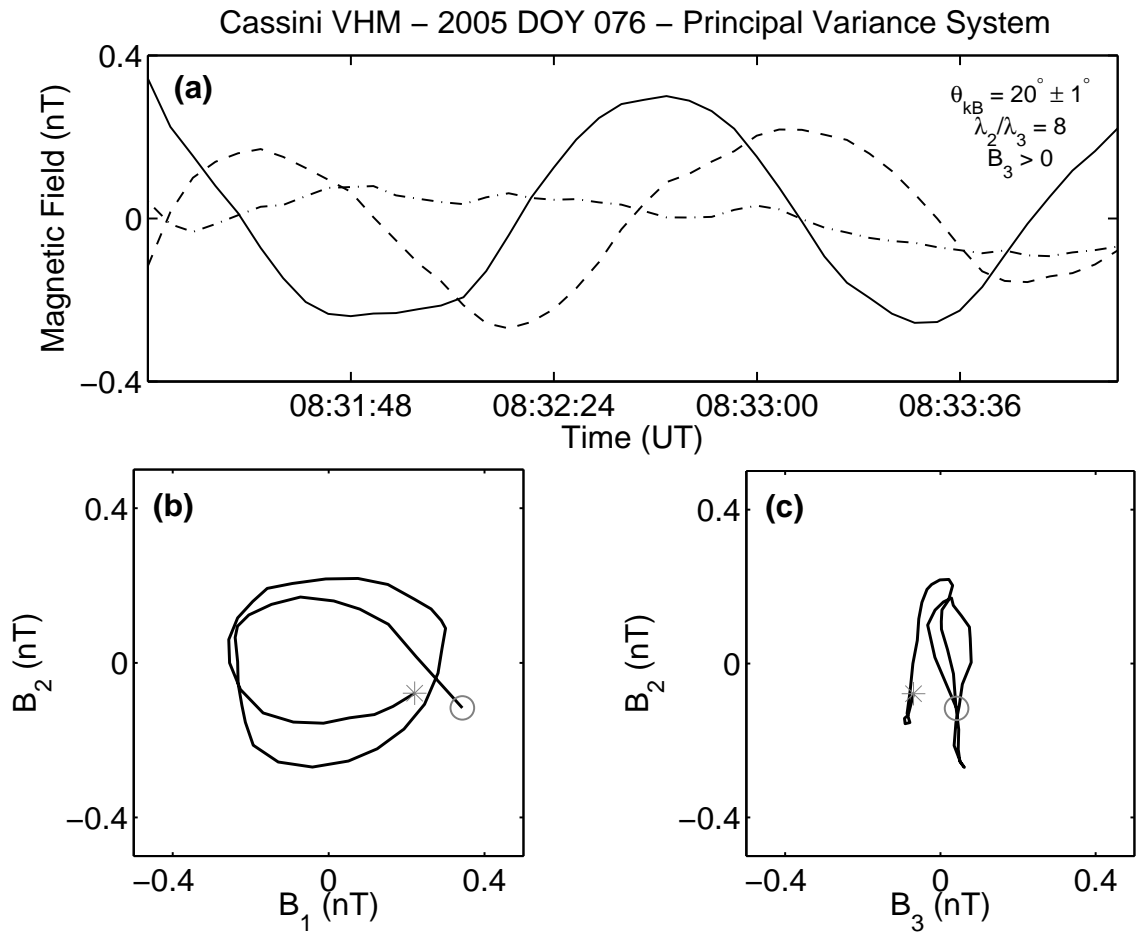}
\end{center}
\caption{(a) Magnetic field components along the maximum (solid line), intermediate (dashed line) and minimum variance direction (dot-dashed line). (b) Hodogram showing the magnetic field in the intermediate-minimum variance plane and (c) in the maximum-intermediate variance plane. The circle and the asterisk indicate the beginning and the end of the hodogram, respectively. }
\label{fig:mva1}
\end{figure}

Figure \ref{fig:waves_m} shows an example of waves between 08:31:08 UT and 10:35:00 UT on 17 March (day 076) 2005. We obtained that these waves are quasi-monochromatic and steepened with frequencies above $\Omega_{\text{H+}}$. In the spacecraft frame, these waves have periods of the order of 60 s. According to the ambient magnetic field magnitude ($\sim$0.3 nT), the period is significantly lower than the local proton cyclotron period  T$_{H+}\sim300$ s. Figure \ref{fig:mva1} shows the MVA results for a quasi-monochromatic wave packet shown in Figure \ref{fig:waves_m} (interval 08:31:12-08:34:04 UT). Figure \ref{fig:mva1}a shows the components of the magnetic field along the maximum, intermediate and minimum variance direction. Figure \ref{fig:mva1}b and \ref{fig:mva1}c show the projection of the wave magnetic field (hodograms) on the maximum-intermediate and the minimum-intermediate variance planes, respectively. The circle and the asterisk indicate the beginning and the end of the hodogram, respectively. A 
large ratio between the intermediate and minimum eigenvalues of the variance matrix ($\lambda_2/\lambda_3\approx10$) shows a clear minimum variance direction. The hodogram in \ref{fig:mva1}c shows that the polarization on the minimum variance plane is circular right-handed with respect to the mean magnetic field (B$_3>0$). The angle between the mean magnetic field and the minimum variance eigenvector is $\theta_{\text{kB}}=20^{\circ}\pm1^{\circ}$, revealing that these waves propagate in a slightly oblique direction with respect to the ambient magnetic field. 

\begin{figure}[h!]
\begin{center}
\includegraphics[width=1\textwidth]{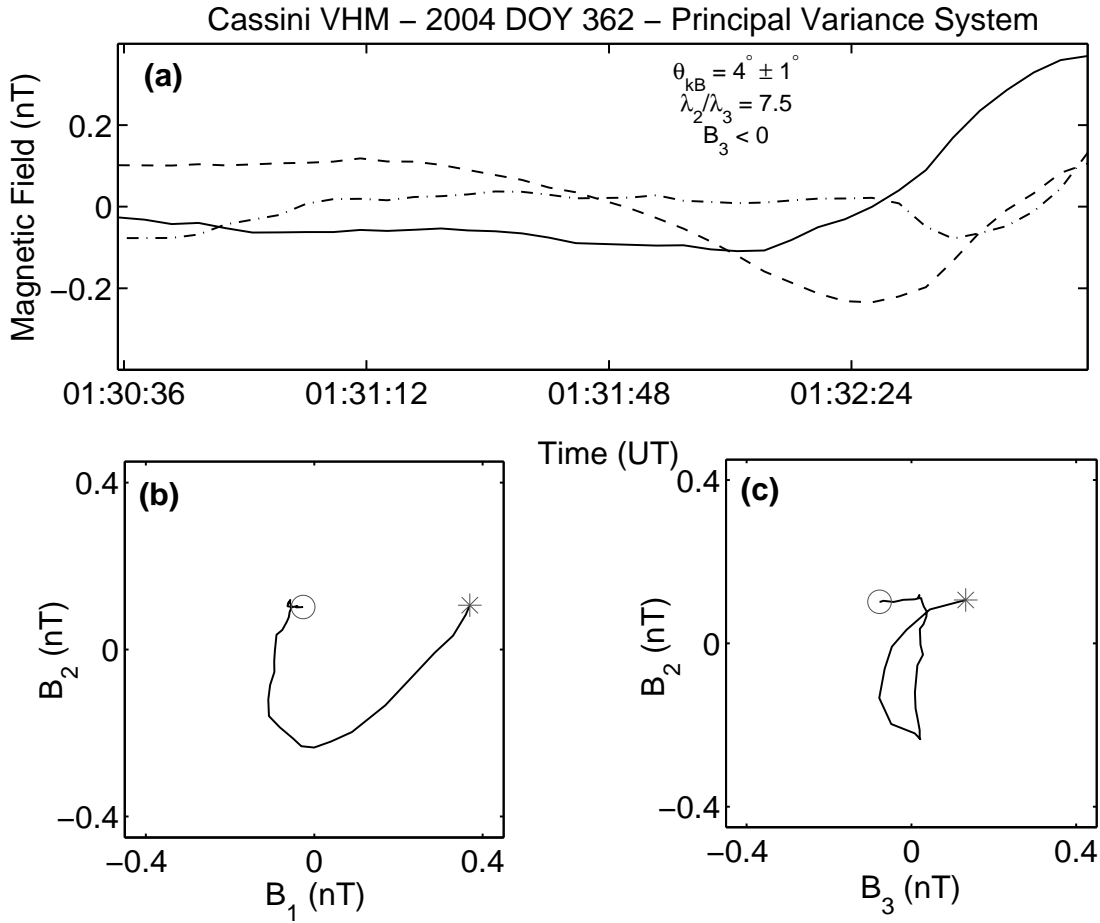}
\end{center}
\caption{(a) Magnetic field components along the maximum (solid line), intermediate (dashed line) and minimum variance direction (dot-dashed line). (b) Hodogram showing the magnetic field in the intermediate-minimum variance plane and (c) in the maximum-intermediate variance plane. The circle and the asterisk indicate the beginning and the end of the hodogram, respectively.}
\label{fig:mva2}
\end{figure}

Figure \ref{fig:waves_nl} shows an example of a wave train seen by Cassini VHM between 01:00:00 UT and 03:00:00 UT on 27 December (day 362) 2004. We obtained that this kind of waves are phase steepened with frequencies smaller than $\Omega_{\text{H+}}$. These waves have periods of the order of 5 to 10 min in the spacecraft frame, and they were the most frequently observed. According to the ambient magnetic field, 2 to 3 times the local proton cyclotron period for an IMF magnitude between 0.35 nT and 0.5 nT. In several cases, we saw a steepening front located at the right of waves with a higher frequency wave packet attached to it. We note a decrease in amplitude and in the period of these oscillations with increasing distance from the steepening front. Figure \ref{fig:mva2} shows the results of the MVA applied on one period of these waves between 01:30:36 UT and 01:33:00 UT. From left to right, it can be seen how an early linear polarization is followed by a circular polarization toward the end of the 
interval. The angle between the minimum variance vector and the mean magnetic field suggests that the propagation of these waves is quasi-parallel ($\theta_{\text{kB}}=4^{\circ}\pm1^{\circ}$) to the ambient magnetic field. The hodogram in Figure \ref{fig:mva2} shows that the magnetic field rotation around the minimum variance direction is left-handed with respect to the ambient magnetic field (B$_3<$0). The circle and the asterisk indicate the beginning and the end of the hodogram, respectively. It is worth noticing that we identified the same two categories than \citet{B2007} had found, for a much bigger Cassini MAG's data set.

\section{Determination of Saturn's ULF wave foreshock boundary: Results}\label{ulf-bound}

\subsection{Data Selection Criterion}\label{dtc}

To identify the ULF wave foreshock boundary using the data set described in section \ref{obs}, we looked for beginnings (or endings) of intervals in which the magnetometer detected ULF waves. Due to the lack of measurements of other physical variables (solar wind density, velocity, dynamic pressure and composition), it is essential to establish a criterion to differentiate a ULF wave foreshock boundary crossing from other possible phenomena, such as discontinuities propagating in the solar wind. 
\begin{figure}[h!]
\begin{center}
\includegraphics[width=1\textwidth]{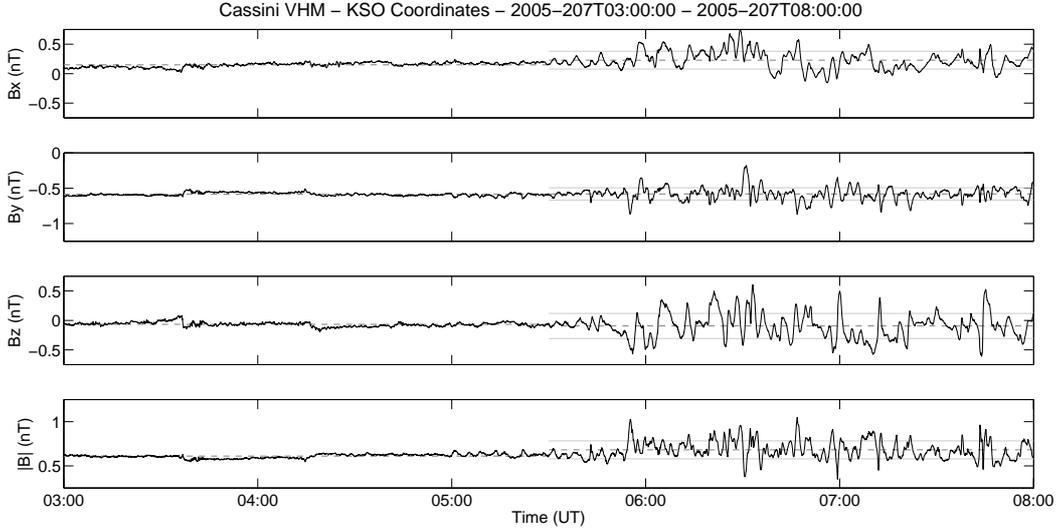}
\end{center}
\caption{Example of stationary crossing (entry) of the ULF wave foreshock boundary as detected by the VHM magnetometer on board Cassini between 03:00:00 UT and 08:00:00 UT on 26 July (day 207) 2005. Average values in the wave zone (${B}^{\text{w}}_\text{j}$) and in the zone without waves (${B}^{\text{nw}}_\text{j}$) are in dashed-gray line. The solid-gray lines correspond to (${B}^{\text{w}}_\text{j}\pm\sigma^{\text{w}}_\text{j}$).}
\label{fig:good_one}
\end{figure}

For a correct determination of the ULF wave foreshock boundary, we considered those crossings in which Cassini was entering or leaving the wave region under steady IMF conditions. Using the whole orbital data set of Cassini's VHM, we identified a total of 59 beginnings or endings of intervals in which the magnetometer detected ULF waves. For this determination, we made no distinction regarding the kind of the field oscillations, requiring only that the transition from or to the wave region would be clearly apparent. We defined the following selection criterion for each component of the magnetic field ($j=x,y,z$): if the difference between the average values in the wave zone (${B}^{\text{w}}_\text{j}$) and in the zone with no waves (${B}^{\text{nw}}_\text{j}$) is smaller than the standard deviation in the wave zone ($\sigma^{\text{w}}_\text{j}$), we consider that Cassini crossed a stationary ULF wave foreshock boundary (Figure \ref{fig:good_one}).
\begin{figure}[h!]
\begin{center}
\includegraphics[width=1\textwidth]{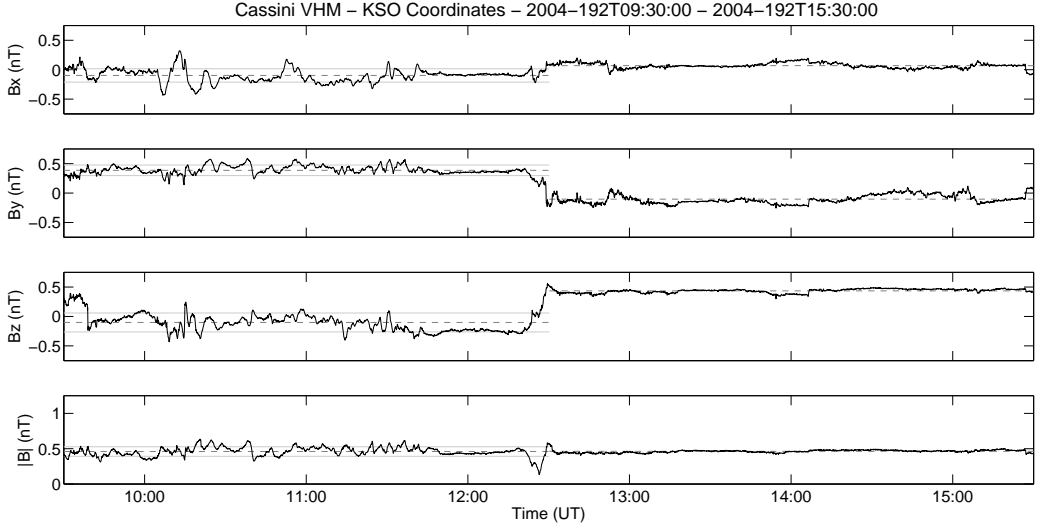}
\end{center}
\caption{Example of non stationary crossing (exit) of the ULF wave foreshock boundary, as detected by the VHM magnetometer on board Cassini between 09:30:00 UT and 15:30:00 UT on 10 July (day 192) 2004. Average values in the wave zone (${B}^{\text{w}}_\text{j}$) and in the zone without waves (${B}^{\text{nw}}_\text{j}$) are in dashed-gray line. The solid-gray lines correspond to (${B}^{\text{w}}_\text{j}\pm\sigma^{\text{w}}_\text{j}$).}
\label{fig:bad_one}
\end{figure}
If any of the three components did not satisfy this condition, the event discarded from our analysis (Figure \ref{fig:bad_one}).

The average values for each component of the magnetic field in the wave zone and in the zone with no waves are the mean values of the observations over two hours before (and after) each apparent crossing. We found that this particular time interval is sufficiently long to obtain values representative of the mean magnetic field, because the longest wave periods found are 10 minutes long. Following this selective criterion, we reduced the original data set of 59 crossings to only 21 stationary crossings. In a second part of the study, for each component of the magnetic field we changed this 1.$\sigma^{\text{w}}_\text{j}$ condition, requesting the difference between ${B}^{\text{w}}_\text{j}$ and ${B}^{\text{nw}}_\text{j}$ to be smaller than 1/2$.\sigma^{\text{w}}_\text{j}$ (more restrictive criterion) and 3/2$.\sigma^{\text{w}}_\text{j}$ (less restrictive criterion).

\subsection{Foreshock Coordinates}\label{res}

In order to identify the ULF wave foreshock boundary independently from the changes in the IMF and the corresponding location of the bow shock, we analyze our data in a particular set of coordinates. In the case of the Earth, \citet{GB1986} introduced the so-called \textit{solar foreshock coordinates}. In this coordinate system, the {\boldmath$x$} axis points toward the Sun (it is parallel to $\hat{\textbf{x}}_\text{kso}$), and the {\boldmath$x$}-{\boldmath$y$} plane is the $\textbf{v}_{\text{sw}}$-$\textbf{B}$ plane which passes through the location of the spacecraft at a particular stationary crossing.

For each crossing, in or out of the ULF wave foreshock under steady IMF conditions, we use the bow shock hyperboloidal fit from \citet{S1985}. We used the eccentricity $e= 1.05\pm0.09$ obtained by \citet{M2008} and kept it constant throughout our study. In each crossing, we estimate the value of $L$ using equation \eqref{conic} and the location of the nearest bow shock crossing. For this purpose, we used the list of times and locations of Cassini's bow shock crossings between 27 June (day 027) 2004 and 12 August (day 224) 2005 published by \citet{M2008}. The error in the determination of the parameter size was $\Delta L= 2~R_S$.

\begin{figure}[h!]
\begin{center}
\includegraphics[width=1\textwidth]{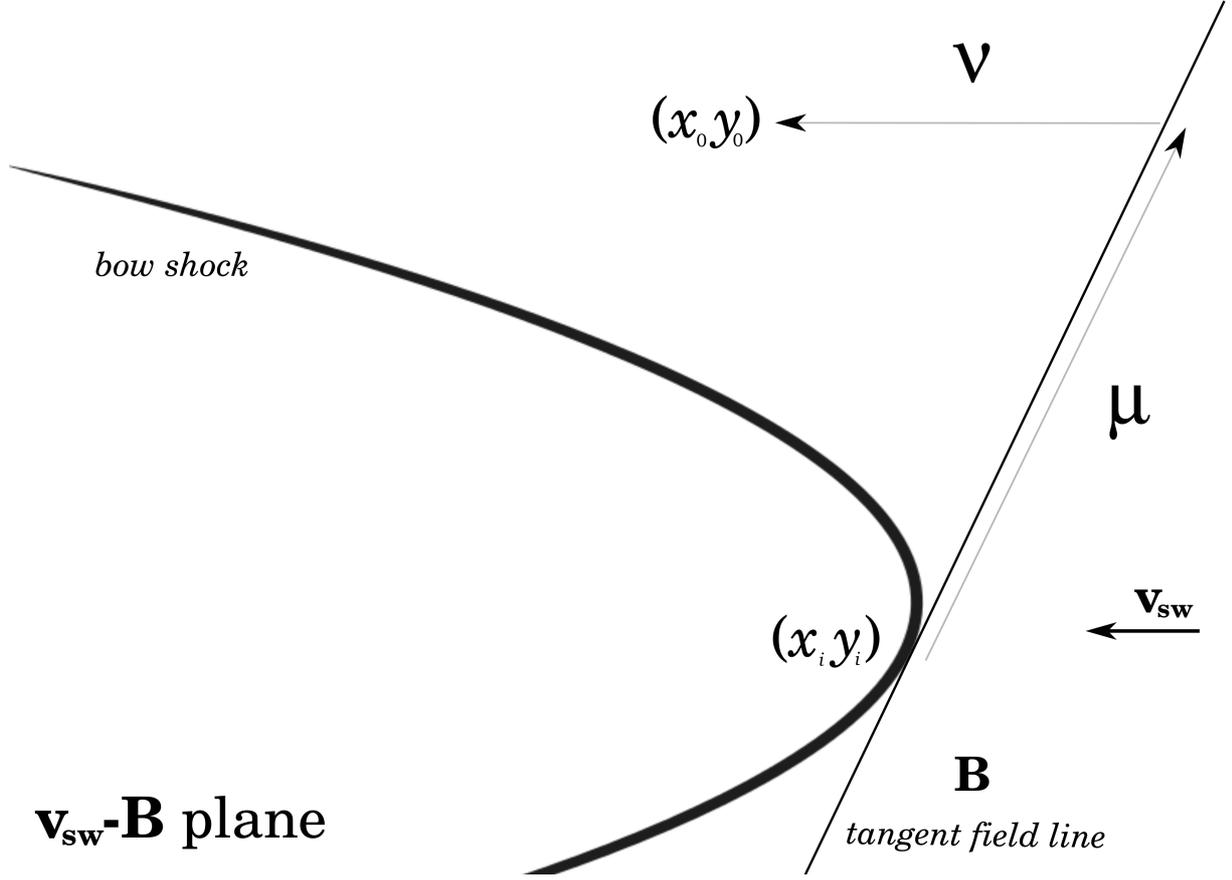}
\end{center}
\caption{Definition of the foreshock coordinates in the $\textbf{v}_{\text{sw}}$-$\textbf{B}$ plane, i.e. the {\boldmath$x$}-{\boldmath$y$} plane. The coordinate $\mu$ is the distance along the tangent magnetic field line between the tangent point and the observation point. The coordinate $\nu$ is the distance along the $\hat{\textbf{x}}_{\text{kso}}$ direction between the tangent magnetic field line and the observation point.}
\label{fig:gb}
\end{figure}

Assuming $\textbf{v}_{\text{sw}}\parallel\hat{\textbf{x}}_{\text{kso}}$, we calculate the foreshock coordinates in the $\textbf{v}_{\text{sw}}$-$\textbf{B}$ plane or {\boldmath$x$}-{\boldmath$y$} plane (see Figure \ref{fig:gb}) as:
\begin{equation}\label{SFC}
\begin{aligned}
\mu &= \frac{(y_o-y_{i})}{\text{sen}\theta_{\text{Bx}}} \\
\nu &= \frac{(y_o-y_{i})}{\text{tan}\theta_{\text{Bx}}} + x_{i} -x_o  
\end{aligned}
\end{equation}
where $\theta_{\text{Bx}} = \text{acos}^{-1}(\textbf{B}\cdot\textbf{x}_{\text{kso}}/\text{B})$ is the IMF cone angle, ($x_i,y_i$) is the intersection point between the tangent IMF line and the bow shock fit, while ($x_o,y_o$) is the observation point (Cassini's crossing location). It is worth mentioning that the $\textbf{v}_{\text{sw}}$-$\textbf{B}$ plane in general does not contain the $\hat{\textbf{x}}_{\text{kso}}$ axis (except for the particular $\textbf{v}_{\text{sw}}$-$\textbf{B}$ plane which passes through the center of the planet). In fact, there should be a third coordinate to measure the distance of any $\textbf{v}_{\text{sw}}$-$\textbf{B}$ plane to the center of the planet. \citet{GB1986} have deliberately ignored that third coordinate, and showed a remarkable correlation in the $\mu$-$\nu$ plane. As shown in Figure \ref{fig:gb}, the non-orthogonal coordinate set ($\mu,\nu$) has its origin in the tangent point ($x_i,y_i$), $\mu$ being the distance along the tangent IMF line and $\nu$ being the 
horizontal distance from that particular field line to 
Cassini's location.

\begin{figure}[h!]
\begin{center}
\includegraphics[width=1\textwidth]{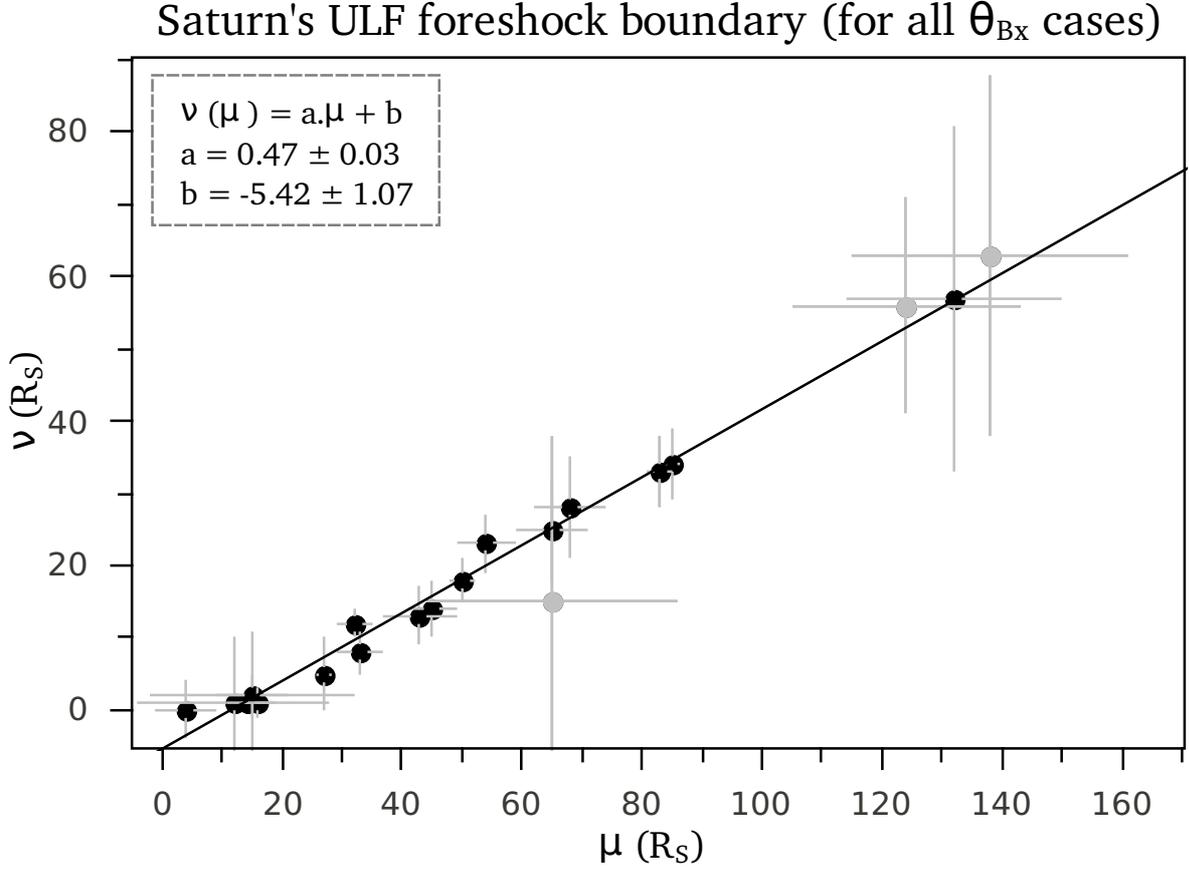}
\end{center}
\caption{ULF wave foreshock boundary for $\theta_{Bx}>45^{o}$ (dark circles) and $\theta_{Bx}<45^{o}$ (gray squares). If we only consider the $\theta_{Bx}>45^{o}$ cases, our best fit yields $\nu(\mu)=a.\mu+b$ ($a=0.47\pm0.04$; $b=-5.65\pm1.36$), which is indistinguishable from the best fit displayed in this Figure.}
\label{fig:sigma1}
\end{figure}

For the 21 stationary crossings considered in section \ref{dtc}, we have calculated their locations in terms of the coordinate set ($\mu,\nu$) . At Earth, the location of the ULF wave foreshock boundary was found to depend on the IMF cone angle {(\citet{GB1986}, \citet{LR1992a})}. For this reason, we analyzed two separate sets of data, one with $\theta_{\text{Bx}}<45^{o}$ and the other with $\theta_{\text{Bx}}>45^{o}$, i.e. for small and large cone angles. We performed a scatter plot of the 21 ULF wave foreshock boundary crossings considered. Figure \ref{fig:sigma1} shows the cases with $\theta_{Bx}>45^{o}$ in black circles and those with $\theta_{Bx}<45^{o}$ in gray circles, and the straight line is our best linear fit (black line) for all our crossings. The small cone angle cases (gray circles) correspond to tangent lines close to the asymptote of the hyperbola ($\theta_{\infty}=\text{cos}^{-1}(1/e)\approx18^{o}$), and therefore have relatively large error bars. However, our best fit considering only the $\
theta_{Bx}>45^{o}$ cases, yields $\nu(\mu)=a.\mu+b$ ($a=0.47\pm0.04$; $b=-5.39\pm1.10$), which is indistinguishable from the result displayed in Figure \ref{fig:sigma1}.

We also studied the variability of this result as our selection criterion become more stringent (see sub-section \ref{dtc}). In Figure \ref{fig:sigma1} we only consider those crossings in which the three components of the magnetic field ($j=x,y,z$) satisfy that the difference between the average values in the wave zone and in the zone without waves is smaller than the standard deviation in the wave zone (1.$\sigma^{\text{w}}_\text{j}$). As we mentioned in sub-section \ref{dtc}, if any of the three components did not satisfy this condition, the crossing was discarded from our analysis.

In a more restrictive criterion, for each component of the magnetic field, we changed the 1.$\sigma^{\text{w}}_\text{j}$ condition to 1/2$.\sigma^{\text{w}}_\text{j}$. In this way we reduced the original data set of 21 stationary crossings to only 16 crossings. Considering only these 16 cases, our best fit yields $\nu(\mu)=a.\mu+b$ ($a=0.46\pm0.04$; $b=-5.26\pm1.12$), which is almost identical to the result displayed in Figure \ref{fig:sigma1}. We also tried with a less restrictive criterion, changing the original 1$.\sigma^{\text{w}}_\text{j}$ to 3/2$.\sigma^{\text{w}}_\text{j}$. In this case our original data set of 21 crossings, increase to 29 crossings, and our best fit yields $\nu(\mu)=a.\mu+b$ ($a=0.49\pm0.03$; $b=-5.99\pm1.01$), which is again, within the errors, the same result displayed in Figure \ref{fig:sigma1}. 
\begin{figure}[h!]
\begin{center}
\includegraphics[width=1\textwidth]{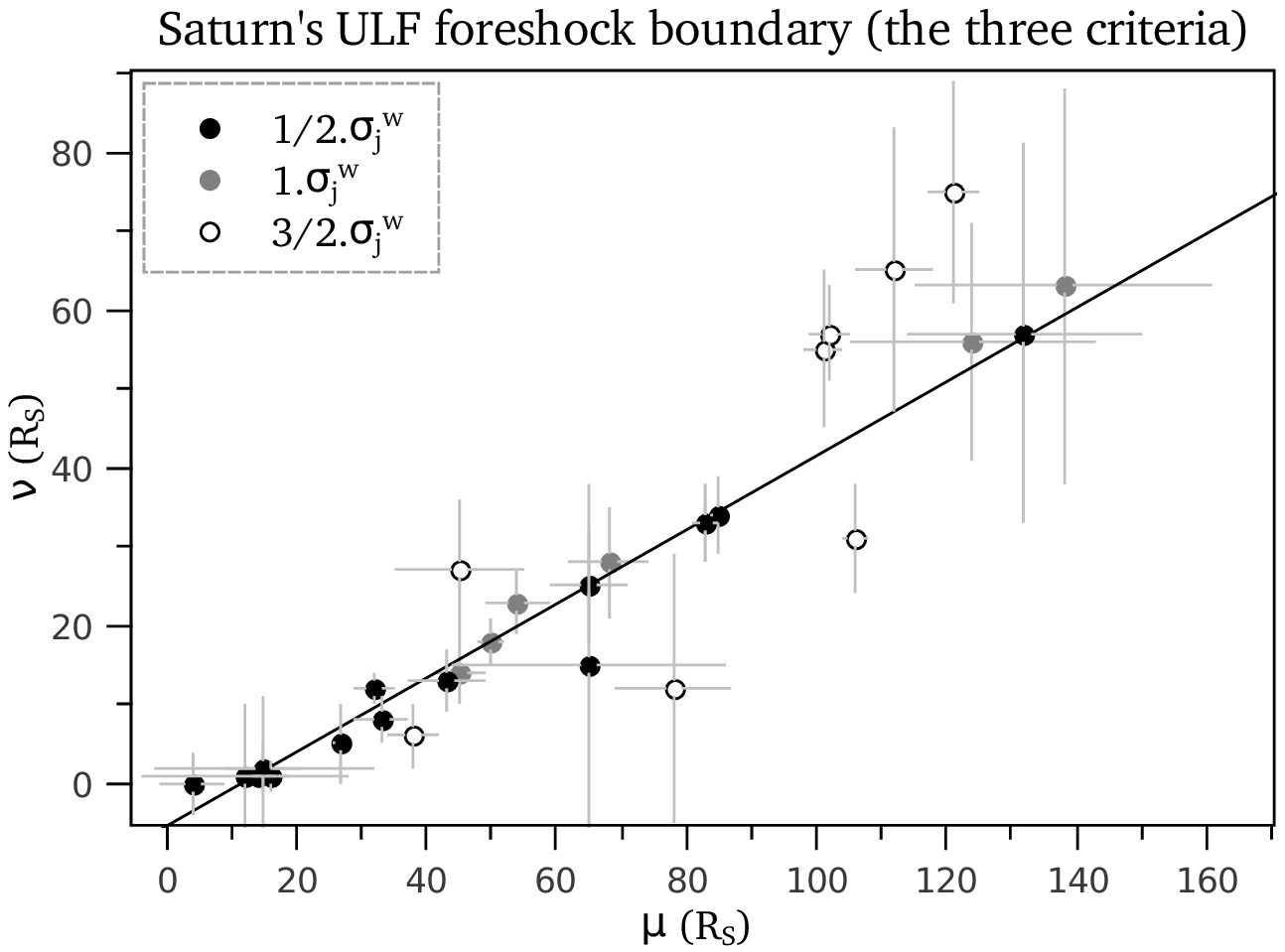}
\end{center}
\caption{ULF wave foreshock boundary for different criteria: the crossings in black correspond to the more restrictive criterion (1/2$.\sigma^{\text{w}}_\text{j}$ condition); the crossings in gray are added when the 1$.\sigma^{\text{w}}_\text{j}$ criterion is considered, while the crossings indicated by white are added when the less restrictive criterion of 3/2$.\sigma^{\text{w}}_\text{j}$ is considered.}
\label{fig:nu_mu_all}
\end{figure}

In Figure \ref{fig:nu_mu_all} we show the results for the three criteria. Black circles are those crossings satisfying the most stringent criterion (1/2.$\sigma^{\text{w}}_\text{j}$), gray circles correspond to crossings between 1/2$.\sigma^{\text{w}}_\text{j}$ and 1$.\sigma^{\text{w}}_\text{j}$, while white circles are those between 1$.\sigma^{\text{w}}_\text{j}$ and 3/2$.\sigma^{\text{w}}_\text{j}$. In all these cases, the uncertainties in the determination of the three magnetic field components were dominated by the statistical error given by the corresponding standard deviation. As mentioned, for the semilatus rectum (size parameter) $L$ and the eccentricity $e$ from the hyperboloidal model, we consider the uncertainties presented in section \ref{ulf-waves}. 

\subsection{$\theta_{Bn}=45^\circ$ Curves}\label{theta}

For each crossing studied, we extend the straight magnetic field line and check for connection to the bow shock fit. As mentioned in the previous section, we checked for magnetic field line connectivity to the bow shock by using the \citet{S1985} fit, with fixed eccentricity $e=1.05\pm0.09$. We varied $L$ so as to make the bow shock fit to coincide with the closest bow shock crossing made by Cassini.

If we find that Cassini is connected to the shock by straight magnetic field lines, we identify the intersection point on the bow shock fit. Once this point has been identified, the angle between the magnetic field and the shock normal at the shock surface, $\theta_{Bn}$, can be calculated. In all the wave events that we identified we found that the magnetic field line intersects the bow shock fit, i.e. Cassini was in the region magnetically connected to the bow shock. In particular, we find that the increment in wave activity is associated to foreshock regions for which $\theta_{Bn}<45^\circ$. In fact, 12 out of the 21 crossings have $\theta_{Bn}=45^\circ\pm5^\circ$, while the remaining 9 crossings have $\theta_{Bn}$ within the range between 35$^\circ$ and 55$^\circ$. If we consider the 21 crossings, the average value is $\theta_{Bn}=42^\circ$ and the standard deviation is $\sigma_{\theta_{Bn}}=9^\circ$.

\begin{figure}[ht!]
{\centering
\includegraphics[width=1\textwidth]{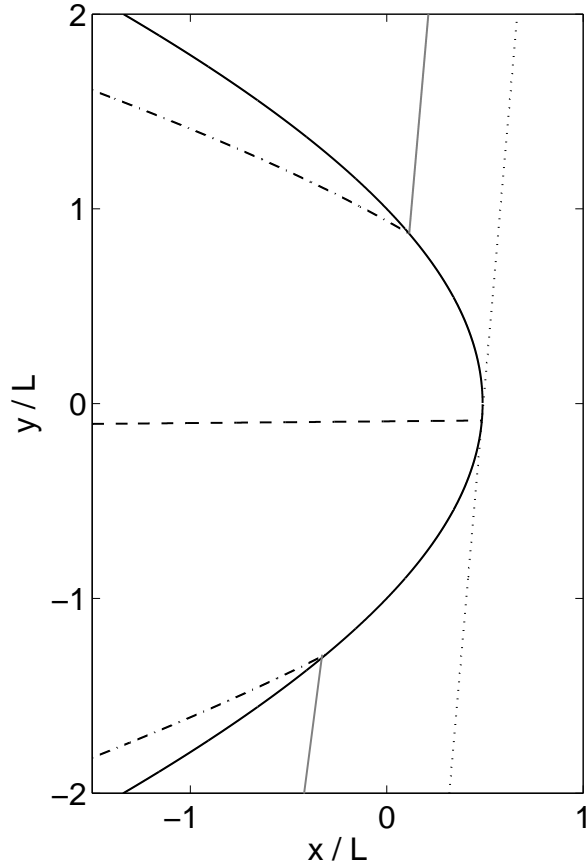}
}
\caption{The curves of $\theta_{Bn}=45^\circ$ (dot-dashed line) and $\theta_{Bn}=90^\circ$ (dashed line). In the $\textbf{v}_{\text{sw}}$-$\textbf{B}$ plane $z=0$, the average bow shock fit (in units of $L$), the tangent field line for $\theta_{Bx}=85^\circ$ (point line), and the magnetic field line corresponding to $\theta_{Bn}=45^\circ$ (solid line).}
\label{fig:curves}
\end{figure}

It is important to emphasize that $\theta_{Bn}=45^\circ$ constitutes a natural and conventional distinction between quasi-parallel and quasi perpendicular shocks. For a given magnetic field orientation and parameter size, there are two lines on the hyperboloidal shock surface, for which $\theta_{Bn}=45^\circ$. In Figure \ref{fig:curves} we show a a schematic view of the bow shock (in units of $L$) and the lines corresponding to $\theta_{Bn}=45^\circ$ (dot-dashed lines) and $\theta_{Bn}=90^\circ$ (dashed lines), where the field lines are tangent to the hyperboloidal shock. The \textbf{x}-\textbf{y} plane in Figure \ref{fig:curves} is parallel to the magnetic field lines and contains the planet in $(x,y)=(0,0)$. Within this context, we claim that the set of magnetic field lines that cross the shock on the $\theta_{Bn}=45^\circ$ curves will determine the ULF wave foreshock boundary. On other hand, a perhaps more abstract identification of the ULF wave foreshock boundary is given by the best fit of crossings on 
the $\mu$-$\nu$ plane \citep{GB1986}.
\begin{figure}[ht!]
\begin{center}
\includegraphics[width=1\textwidth]{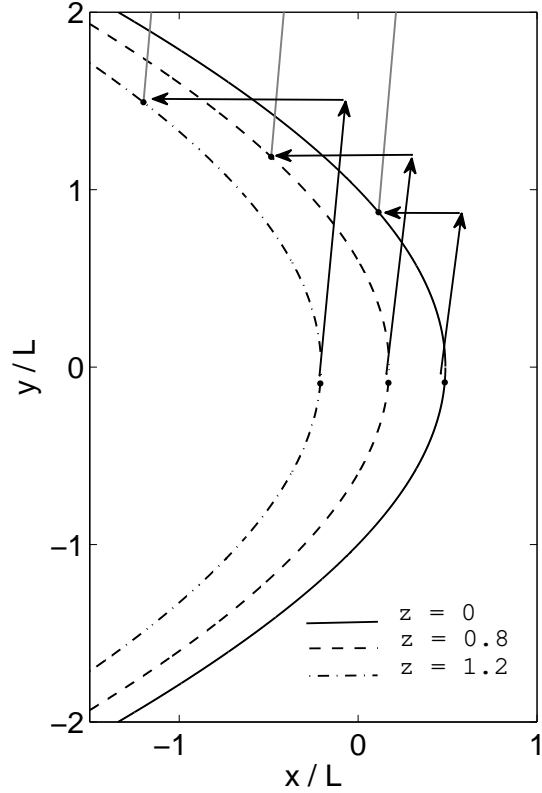}
\end{center}
\caption{The ($\mu$,$\nu$) pair for the location of points satisfying $\theta_{Bn}=45^\circ$ at different (parallel) $\textbf{v}_{\text{sw}}$-$\textbf{B}$ planes for $\theta_{Bx}=85^\circ$ and parameter size $L=$ 51 $R_S$. The distance of each plane to the planet is labeled in units of $L$.}
\label{fig:trian}
\end{figure}

Therefore, we decided to study the consistency of our linear best fit in the $\mu$-$\nu$ plane and the curve corresponding to $\theta_{Bn}=45^\circ$. Assuming a constant $\theta_{Bx}$ (the magnetic field lines lay in the {$\textbf{x}$}-{$\textbf{y}$} plane) and parameter size $L=$ 51 $R_S$, we computed the ($\mu$,$\nu$) pair for the location of the point that satisfies $\theta_{Bn}=45^\circ$ at different (parallel) $\textbf{v}_{\text{sw}}$-$\textbf{B}$ planes (see Figure \ref{fig:trian}). The corresponding distance of each plane (in units of $L$) to the planet is labeled. We can clearly see in Figure \ref{fig:trian} an approximate proportionality between the $\mu$-$\nu$ coordinates at any given plane. This implies that if Cassini crossed right at the intersection between the bow shock and the $\theta_{Bn}=45^\circ$ field line, we should expect a straight line passing through the origin in the $\mu$-$\nu$ plane (see Figure \ref{fig:sigma1}). The fact that Cassini crosses that very field line necessarily at a 
distance from that intersection toward the upstream direction, causes a systematic increase in the value of $\mu$, leaving $\nu$ essentially unchanged. The distance from the bow shock measured along field lines range between $0.1~R_S$ and $17~R_S$ for the 21 crossings considered. On the one hand, this effect will reduce the quality of the linear fit, and on the other hand it will shift the line on the $\mu$-$\nu$ plane to the right (negative intercept), exactly as observed in Figure \ref{fig:sigma1}.

\section{Discussion}\label{dis}

All the wave events detected by Cassini's magnetometer show evidence of magnetic connectivity to Saturn's bow shock, which is a clear indication that these waves are associated with Saturn's foreshock. In absence of a reliable physical model of the bow shock, we used the empirical model of \citet{M2008}, which is based on extensive observations from Cassini made during the same time span of our work (15 months from June 2004 through August 2005). Assuming that the bow shock is axially symmetric about the solar wind flow direction, the average bow shock location is obtained by fitting a conic section to the bow shock crossings using a nonlinear least squares technique. This fit provides an adequate description of the average shape, given by an eccentricity of $e=1.05\pm0.09$, and also that of its average location, corresponding to $L=(51\pm2)~R_S$. In our study, we assume that the shape of Saturn's bow shock remains constant throughout our observational time span, while its location varies in response to time 
variations of the solar wind's ram pressure. In other words, for each wave event or ULF wave foreshock boundary crossing, we assumed $e$ to remain constant. Meanwhile, the semilatus rectum $L$, which determines the distance of the bow shock to the planet, was allowed to change to account for this pressure effect. We found values of $L$ between $53\pm2~R_{S}$ and $78\pm2~R_{S}$. We also calculated the parameter size using an average between the two closest bow shock crossing locations for a given event (wave or crossing). For the sake of comparison, we also made our calculation assuming that the parameter size remains constant ($L= 62~R_S$, which is the average value) for all the wave events. We found, that the bow shock fit was not representative of many of the upstream events studied, since they fell in the downstream region of this static fit. Therefore, we concluded that the bow shock fit calculated using equation \eqref{conic} with a varying $L$ parameter gives a more accurate empirical representation of 
the real Kronian bow shock.

For all the wave events observed, we performed a preliminary characterization based on their frequency and polarization as measured in the spacecraft reference frame. According to their frequencies, we observe two different types of oscillations. Waves with frequencies below $\Omega_{\text{H+}}$ are the most frequently observed, they are phase steepened with periods of the order of 5 to 10 minutes in the spacecraft frame. As we mentioned in section \ref{ulf-waves}, we saw steepening fronts located at the right of waves with a higher frequency wave packets attached to them. We also note a decrease in amplitude and in the period of the oscillations within each wave packet with increasing distance from the steepening front. The angle between the minimum variance vector and the mean magnetic field suggests that the propagation of these waves is quasi-parallel. The magnetic field rotation around the minimum variance direction is left-handed with respect to the IMF direction. As was discussed in \citet{B2007}, 
these signatures suggest that these waves are ion/ion resonant right-hand (fast magnetosonic) mode waves which steepen during the nonlinear regime and emit a dispersive whistler to stop the steepening. We also observe waves with frequencies above $\Omega_{\text{H+}}$, which are either quasi-monochromatic or steepened, with periods of the order of 60 s (significantly lower than the local proton cyclotron period  T$_{\text{H}+}\sim300$ s). For the quasi-monochromatic events, their corresponding hodograms show that the polarization is circular right-handed with respect to the mean magnetic field. The angle between the ambient magnetic field and the minimum variance direction reveals that these waves propagate in a slightly oblique direction with respect to the IMF orientation. This kind of wave packets is an exception rather than the rule (we found ten events only). At the Earth's foreshock, there is a particular type of ULF waves, the so-called 30 s (period) waves which are quasi-monochromatic \citep[e.g.][]{
LR1992b}. This kind of waves are found always near to the ULF wave foreshock boundary. In our study we did not find quasi-monochromatic waves next to the Saturnian ULF wave foreshock boundary. A more detailed statistical study would shed some light about the nature of this second group of waves and whether they are the Saturnian equivalent of the 30 s modes found in the Earth's foreshock.
 
When Cassini is magnetically connected to the bow shock fit by straight magnetic field lines (since we are assuming a uniform external magnetic field), we identify the intersection point between the field line and the bow shock fit as well as the angle $\theta_{Bn}$. In all the crossings that we observed, we find that the presence of magnetic fluctuations is associated with values of $\theta_{Bn}$ corresponding to quasi-parallel geometries ($\theta_{Bn}<45^\circ$). This result is fully consistent with the one obtained by \citet{LR1992a} at Earth. Using ISEE data, they examined the presence of ULF waves in the region immediately upstream from the Earth's bow shock. Their statistical study shows a well-defined boundary on the $\textbf{v}_\text{sw}$-\textbf{B} plane, separating the disturbed from the undisturbed magnetic field at $\theta_{Bn}\sim50^\circ$. Our result for Saturn's foreshock is consistent with this scenario, regardless of whether these waves are immediately adjacent to the bow shock or deep into 
the foreshock region. \citet{S1983} developed a theoretical framework for studying the general trajectories of ions reflected or leaked upstream from the Earth's bow shock and subjected only to the Lorenz force in a steady IMF. In the case known as specular reflection, the ions bounce off the shock potential reversing their velocity component along the shock normal $\hat{n}$, while maintaining their parallel component. In the de Hoffman-Teller frame \citep{HT1950}, assuming that the bow shock is infinitesimally thin and has no overshoot, \citet{S1983} found that after interacting with the shock, the guiding center motion of incident solar wind ions is reflected to the upstream region with $\theta_{Bn}<45^\circ$. For $\theta_{Bn}>45^\circ$, specularly reflected ions always re-encounter the bow shock with sufficient energy to penetrate into the downstream region. It is worth mentioning that these specularly reflected ions oriented to the upstream region for quasi-parallel geometries would have no conserving 
magnetic moment, in contrast with the field-aligned reflected beams, for which there is a magnetic moment conservation \citep{P1980}. In the absence of Cassini plasma observations that could confirm the presence of backstreaming ions giving rise to an ion foreshock region, we cannot observationally associate the presence of waves with specific ion distributions. However, if the mechanism of generation of these waves is local, i.e. the wave growth rate $\gamma$ is sufficiently large, we could expect specularly reflected ions to be responsible for these waves since they are the only population whose spatial distribution depends on $\theta_{Bn}$. In this case, for quasi-parallel geometries, the reflected ions can go upstream from the bow shock and be detectes through ULF wave activity. Nevertheless, a full discussion on the wave generation caused by particle distributions as a function of $\theta_{Bn}$, is beyond the purpose of this study.

For a correct determination of Saturn's ULF wave foreshock boundary, we only considered crossings in which Cassini was entering or leaving the wave zone under steady IMF conditions, i.e. stationary crossings. We defined a criterion based on the variability of the magnetic field in order to make a proper identification of these stationary crossings. As a result, we reduced our original list of crossings to only 21 stationary ones. At Earth, using single spacecraft techniques, \citet{GB1986} obtained different orientations of ULF wave foreshock boundary for different cone angles (more specifically, they considered two ranges: $\theta_{{Bx}}=20^{\circ}$ - $30^\circ$ and $\theta_{Bx}=40^\circ$ - $50^\circ$). In agreement with Parker's angle prediction for Saturn \citep{J2008}, our study shows that 83\% of the crossings correspond to cone angles larger than 55$^\circ$. Therefore, the range of $\theta_{Bx}$ considered is different to that at Earth. Nevertheless, we analyzed the correlation between $\mu$-$\nu$ 
coordinates for two ranges of $\theta_{Bx}$: one set of data with $\theta_{{Bx}}<45^{\circ}$ and the other with $\theta_{{Bx}}>45^{\circ}$. A comparison of the best linear fit for the two data set indicates no significant difference between them (see Figure \ref{fig:sigma1}). 

This result, however, does not rule out a dependence on $\theta_{{Bx}}$ since the errors in the determination of $\mu$-$\nu$ become larger with decreasing $\theta_{Bx}$. More specifically, the main source of error propagation in the determination of each ($\mu,\nu$) pair is the intersection point between the tangent IMF line and the bow shock fit. This intersection point depends critically on the IMF cone angle, which is therefore the main source of error. In particular, small cone angles correspond to almost tangent magnetic field lines close to the asymptotes of the hyperbola. On the other hand, the appearance of waves is observed to occur at $\theta_{Bn}$ values of $\sim45^{\circ}$. The surface determined by the condition $\theta_{Bn}=45^\circ$ is mapped on the $\mu$-$\nu$ plane as straight line, as illustrated by Figure \ref{fig:trian}. As a result, we find that there is an equivalence between the ULF wave foreshock boundary determined by $\theta_{Bn}\sim45^{o}$ and the best fit in the $\mu$-$\nu$ plane \
citep{GB1986}. They are in fact two alternative ways to determine the same spatial region within the foreshock.

Finally, we studied the robustness of the $\mu$-$\nu$ correlation as our selection criterion becomes more stringent. For instance, we change the 1.$\sigma^{\text{w}}_\text{j}$ condition to a more stringent value of 1/2.$\sigma^{\text{w}}_\text{j}$ for the three components of magnetic field, or change it into the less restrictive value of 3/2$.\sigma^{\text{w}}_\text{j}$ for each component. We find that regardless of whether we choose a more or less restrictive criterion, our best fit remains the same, within the fit errors. We also find that for these three different criteria, the boundary does not seem to depend on whether the cone angle is large or small.

\section{Conclusions}\label{con}

Using Cassini's data from SOI in June 2004 through August 2005, we conducted a detailed survey and analysis of ULF waves upstream from Saturn's bow shock. All the wave events show evidence of magnetic connectivity to Saturn's bow shock, therefore we conclude that these wave events are undoubtedly associated with Saturn's foreshock.

As for their frequencies, we identify two distinct types of wave populations. The most frequently observed have frequencies below $\Omega_{\text{H+}}$ and are phase steepened, with periods of the order of 5 to 10 minutes in the spacecraft frame. These waves are often accompanied by precursor whistler wave trains. In agreement with previous works, we suggest that these waves are ion/ion resonant right-hand (fast magnetosonic) mode waves which steepen during the nonlinear regime and emit a dispersive whistler to stop the steepening. We also observe waves with frequencies above $\Omega_{\text{H+}}$, which appear as either quasi-monochromatic or steepened, with periods of $\sim$ 1 minute. The quasi-monochromatic events have a circular right-handed polarization with respect to the mean magnetic field and a propagation slightly oblique with respect to the IMF. A more detailed study will be needed in order to conclude whether the latter are the Saturnian equivalent of the 30 s modes found at Earth. It is worth 
noticing that we have identified the same two categories than \citet{B2007} have found, although for a much bigger Cassini MAG's data set.

We identified 21 stationary crossings inbound and outbound from the ULF wave foreshock region. We calculated their solar foreshock coordinates in the $\textbf{v}_{\text{sw}}$-$\textbf{B}$ plane and we have identified for the first time Saturn's ULF wave foreshock boundary. In the $\mu$-$\nu$ plane we do not find a clear dependence between the foreshock boundary and the IMF cone angle. We also found that the presence of waves is associated with the change in $\theta_{Bn}$ to quasi-parallel geometries. Moreover, we find that our determination of the ULF wave foreshock boundary as the surface given by $\theta_{Bn}=45^\circ$, is indeed consistent with the linear fit in the $\mu$-$\nu$ plane, first proposed by \citet{GB1986} for the Earth's ULF wave foreshock. In this regard, we speculate that the specular reflection is the candidate process for the reflected ions (gyrophase-bunched and diffuse distributions), since it is the only process to produce a signature in $\theta_{Bn}$. Finally, we studied the robustness 
of the $\mu$-$\nu$ correlation for a more restrictive criterion. We found that regardless of the criterion that we use, we obtain the same ULF wave foreshock boundary in the $\mu$-$\nu$ plane.

\end{document}